\documentclass[11pt,a4paper]{article}
\pdfoutput=1
\usepackage{jheppub}
\usepackage{amsmath,amscd}
\usepackage{amsfonts}
\usepackage{amssymb}
\usepackage{graphicx}
\usepackage{color}
\usepackage[normalem]{ulem}
\usepackage{hyperref}
\usepackage{enumerate}

\newcommand{\RR}{\mathbb{R}} 
\newcommand{\ZZ}{\mathbb{Z}} 

\newcommand{\G}{\mathcal{G}}

\def\tr         {{\rm  tr}}
\def\cala         {{\cal A}}

\def\calc         {{\cal C}}

\def\calh         {{\cal H}}

\def\call         {{\cal L}}
\def\calm         {{\cal M}}
\def\caln         {{\cal N}}

\def\be{\begin{equation}}
\def\ee{\end{equation}}
\def\bea{\begin{eqnarray}}
\def\eea{\end{eqnarray}}

\def\a{\alpha}
\def\b{\beta}
\def\h{\eta}
\def\g{\gamma}
\def\G{\Gamma}
\def\d{\delta}

\def\D{\Delta}
\def\l{\lambda}
\def\L{\Lambda}
\def\k{\kappa}
\def\f{\phi}

\def\x{\xi}
\def\n{\nu}
\def\o{\omega}
\def\O{\Omega}
\def\p{\pi}

\def\t{\tau}

\def\sF{{{ F}\!\!\!\!\hskip.8pt\hbox{\raise1pt\hbox{/}}\,}}
\def\som{{{ \omega}\!\!\!\!\hskip.8pt\hbox{\raise1pt\hbox{/}}\,}}
\def\sJ{{{\rm J}\!\!\!\!\hskip.8pt\hbox{\raise1pt\hbox{/}}\,}}

\def\pa{\partial}

\def\to{\rightarrow}
\def\nonu{\nonumber \\{}}
\def\half{{1 \over 2}}

\def\md{{\rm mod\ }}

\def\red{\textcolor{red}}
\def\blue{\textcolor{blue}}

\title{A note on ensemble holography for rational tori}
\author[a]{Joris Raeymaekers}

\affiliation[a]{CEICO, Institute of Physics of the Czech Academy of Sciences,\\  Na Slovance 2, 182 21 Prague 8, Czech Republic.}
\emailAdd{joris@fzu.cz}

\abstract{
	 We study simple examples of ensemble-averaged holography in 
	  free compact boson CFTs with rational values of the  radius squared. These  well-known rational CFTs have an extended chiral algebra generated by three  currents.
	We consider the  modular average of the vacuum character 
	in these theories, which results in  a weighted average over all 
	modular invariants. In the simplest case, when the chiral algebra is primitive (in a sense we explain), 
	the weights in this ensemble average are all equal. In the  non-primitive case the ensemble weights are 
	governed by a semigroup structure on the space of modular invariants. 
	
These observations can be viewed as evidence for a holographic duality between the ensemble of CFTs and an 
	exotic gravity theory based on a compact $U(1) \times U(1)$ Chern-Simons action. 
	In the bulk description, the  extended chiral algebra arises  from soliton sectors, and 
	 	including these  in the path integral on 
	thermal AdS$_3$ leads to the vacuum character of the  chiral algebra. We also comment on wormhole-like contributions to the multi-boundary path integral.
}
\arxivnumber{}
\keywords{}
\begin{document}
 \maketitle
\section{Introduction and summary}
Recent insights into the black hole information puzzle \cite{Penington:2019kki,Almheiri:2019qdq} have
rekindled interest in the interpretation of the sum over topologies in 
quantum gravity.  
At least in  low-dimensional examples, the path integral in a low energy effective gravitational theory has been found
to be  agnostic as to the precise UV completion   and to  compute an ensemble average  over  UV-complete theories. A striking example is that of two-dimensional Jackiw-Teitelboim quantum gravity \cite{Teitelboim:1983uy,Jackiw:1984je}, which was shown  \cite{Saad:2019lba} to compute quantities in  a random ensemble  of quantum mechanical theories.

An early precursor to these insights was Maloney and Witten's \cite{Maloney:2007ud} path integral computation of the partition function of pure AdS$_3$ gravity. The sum over known saddles leads to an average over the modular group of the path integral on thermal AdS$_3$, which is simply the
(Virasoro $\times$ Virasoro)  vacuum character.
 Their result has  features characteristic of an ensemble average over CFTs. 
 Aside from other puzzles \cite{Keller:2014xba,Benjamin:2019stq}  (for recently proposed cures see \cite{Benjamin:2020mfz,Maxfield:2020ale}), the lack of understanding of the moduli space of Virasoro CFTs has so far precluded a  precise ensemble interpretation of their computation. 

A sharper version of averaged AdS$_3$ holography arises when considering theories with extended symmetry in the form of an abelian current algebra \cite{Afkhami-Jeddi:2020ezh,Maloney:2020nni}. In this case, the   space of CFTs with this chiral algebra is a  Narain moduli  space,  and the modular average of the vacuum character
 was indeed  shown to yield an ensemble average over Narain CFTs.   The tentative bulk dual is a non-compact abelian Chern-Simons theory,  treated  as a theory of gravity in the sense that one  sums over topologies in path integral. A justification for this is that the boundary theory automatically includes a stress tensor, which is  a composite of the chiral currents. However, the prescription specifying which topologies are to be included is far from clear.  Various extensions of this example have been studied since then \cite{Perez:2020klz,Dymarsky:2020pzc,Datta:2021ftn,Benjamin:2021wzr,Ashwinkumar:2021kav,Dong:2021wot,Collier:2021rsn,Benjamin:2021ygh}.

While  Narain holography involves an ensemble of irrational  CFTs, the idea of averaged holography also applies to  rational CFTs, which should  provide a simplified  setting to sharpen our understanding of  its  conceptual issues.
In the rational case, the modular average 
 coming from the bulk path integral  reduces  to a finite sum, and can be interpreted as a weighted average over all modular invariants of the relevant chiral algebra, which are also finite in number. Understanding the bulk side of the duality is in these examples  facilitated by the well-studied realization of rational CFTs as Chern-Simons theories with compact gauge group \cite{Witten:1988hf,Moore:1989yh,Elitzur:1989nr}.
 Examples in the literature include the study of  modular averages 
in Virasoro minimal models in \cite{Castro:2011zq}, which  presaged  averaged holography, 
and the  extension to classes of WZW models 
in \cite{Meruliya:2021utr,Meruliya:2021lul}.
 An  open issue in the rational case is that, in contrast to Narain duality where there is a natural measure on moduli space, a physical understanding of (and a general formula for) the weights in the ensemble average is lacking. 
 
In this note we add to this  list of examples by studying averaged holography in a particularly simple setting, which can be thought of as a rational version of Narain duality. 
We consider  the class of `rational torus'  CFTs \cite{Moore:1988ss,Moore:1989yh}, which are   free compact bosons at rational values of the radius-squared,
\be 
R^2 = {p \over q}.\label{rationalR}
\ee
As we review in Section \ref{Secrattor}, these models have an extended chiral algebra which is completely determined by a `level' $k \equiv pq$, and consists of one spin-1    and two spin-$k$ currents. It was shown in \cite{Cappelli:1986hf,Cappelli:1987xt} that the $R^2 = p/q$ compact boson CFTs with the same value of $pq$ form a complete set of modular invariant theories for this chiral algebra. We will study  Poincar\'e sums of the type 
 \be 
Z_{\rm grav}= \caln  \sum_{\g \in \G_f \backslash  PSL(2,\ZZ)}
| \chi_{0,k} (\g \t) |^2 
\label{Poincsumintr}
 \ee
 where $\caln$ is a normalization constant, $\chi_{0,k}(\t )$ is the vacuum character  of the chiral algebra and $\G_f$ is the  subgroup of $PSL(2,\ZZ)$ leaving $| \chi_{0,k}  |^2$ invariant. 
The  quantity (\ref{Poincsumintr}) can be expanded in a basis of modular invariants, and we will interpret the coefficients in this expansion as  ensemble weights in an average over 
rational torus CFTs. A significant part of this note is devoted to computing these ensemble weights. 

The simplest situation occurs when the level $k$ is a square free integer. 
The chiral algebra is then `primitive' in the sense that it is not contained in any larger rational torus algebra. 
 The set of modular invariants forms  in this case  a finite multiplicative group \cite{Cappelli:1986hf}, and we will show in Section \ref{Secprim} that this implies that all the theories in the ensemble appear with equal weights.
   One could argue that these examples provide the simplest infinite class of ensemble averaged theories discussed so far.    
  When the chiral algebra is not primitive, the situation is more complicated as some modular invariants are also invariants of a larger algebra, and the ensemble weights are no longer equal. As we show in Section \ref{Secnonprim},  the  weights are in this case governed by  a semigroup structure on the space of modular invariants.
 
Similar  to the examples in the literature mentioned above, the
 tentative bulk dual of the weighted average over CFTs is  an exotic gravity theory based on a Chern-Simons action. In the case of interest the latter consists of two compact $U(1)$ Chern-Simons actions at  levels $k$ and $-k$. An important insight of \cite{Moore:1989yh,Elitzur:1989nr} is that, in contrast to the non-compact Chern-Simons bulk theories of \cite{Afkhami-Jeddi:2020ezh,Maloney:2020nni}, the  theory on 
 Euclidean AdS$_3$ contains  solitonic sectors, which correspond to winding sectors in the boundary theory. 
The lowest-lying bulk solitons correspond precisely to the two spin-$k$ currents extending the chiral algebra.
 It is an interesting feature of these examples that the extended chiral algebra arises in the bulk from solitons rather than from elementary fields. 
In Section \ref{Secbulk}, we  fill a small gap in the literature by pointing out that the partition function on  Euclidean AdS$_3$, including the sum over the  soliton sectors, 
leads precisely to the vacuum character   $ |\chi_{0,k}|^2$. Boldly assuming the sum over geometries to comprise the $PSL(2, \ZZ )$ family of geometries appearing in semiclassical gravity \cite{Maldacena:1998bw} then leads to the 
 Poincar\'{e} sum in  (\ref{Poincsumintr}).

 We also comment, in section \ref{Secmultibound}, on wormhole-like contributions  to the path integral in the presence of multiple toroidal boundaries. As shown in \cite{Meruliya:2021utr}, these are severely constrained by consistency of the ensemble interpretation. For the primitive rational tori, these constraints simplify significantly and we show that they can be solved  analytically in the class of theories with only two modular invariants.

\section{Rational torus CFTs, Poincar\'{e} sums  and averages}\label{Secrattor}
In this section we collect some results on rational torus CFTs, their chiral algebra, modular invariants and Poincar\'{e} sums of the type (\ref{Poincsumintr}).
\subsection{Chiral algebra}
We start by reviewing (see \cite{Moore:1988ss,Moore:1989yh}  for more details) rational CFTs which possess an extended chiral symmetry algebra  $\cala_k$ characterized by a positive integer `level' $k$.  
The algebra is generated by 
three chiral currents: a spin-1 current $J$  and two spin-$k$ currents  $W^+$ and $W^-$. It is most easily described 
 by its free field realization 
in  terms of a compact free boson  at a  rational value of the radius squared.
 Choosing a pair of relatively prime positive integers $p,q$ 
such that $k = p q$ ($p$ and $q$ are allowed to  be one), 
we  consider a compact boson at radius\footnote{Our conventions for the compact boson CFT follow  Polchinski's book \cite{Polchinski:1998rq} with $\a'=1$.} $R^2 = p/q$. This theory contains primaries of the form
\be 
:e^{i p_L X_L (z)} e^{i p_R X_R (\bar z)}:
\ee 
with 
\be
p_{L,R} =m \sqrt{q\over p}  \pm w  \sqrt{p \over q} ,\qquad m, w \in \ZZ,
\ee
which have dimensions
\be 
\D = {p_L^2 \over 4},\qquad \bar \D= {p_R^2 \over 4}.
\ee
The operators with $\bar \D=0$ in the theory form   the chiral algebra  $\cala_k$. It is generated by  the chiral currents 
\be 
J =  \pa X_L, \qquad W^\pm = :e^{\pm 2 i \sqrt{k } X_L}:\label{freereal}
\ee
These form a nonlinear algebra with mode commutation relations of the form \cite{Moore:1988ss}:
\bea
\, [ J_m, J_n]&=& -{1\over 2} m \d_{m+n,0}\\
\, [ J_m, W^{\pm}_n ] &=& \mp \sqrt{ k}  W^{\pm}_{m+n}\\
\, [W^+_m, W^-_n]&=& { (m^2 -(k-1)^2) (m^2 -(k-2)^2)\ldots (m^2 -1)m \over (2k -1)!} \d_{m+n,0}+ \ldots \nonu
&&+ {\left( 2 i \sqrt{k } \right)^{2k-1} \over (2k-1)!} :J^{2k-1}:_{m+n},
\eea
where in the last line we have omitted terms involving  powers of the current  $J$ lower than $2 k-1$. The  algebra contains a $c=1$ Virasoro subalgebra with stress tensor $T = : J^2:$.

Two remarks will be important for what follows.
A first point is that, from the realization (\ref{freereal}) it follows that, for any positive integer $\a$, the fields $J$ and $:(W^\pm)^\a:$ generate the algebra $\cala_{\a^2 k}$. Therefore $\cala_{\a^2 k}$
	is a subalgebra of $\cala_{k}$ for any $\a$. If $k$ has no quadratic divisors, then $\cala_{k}$ is not contained in any  $\cala_{k'}$ with $k'\neq k$, and we will say that  $\cala_{k}$ is  {\em primitive}. If $k$ does have quadratic divisors we will call  $\cala_{k}$ {\em non-primitive}.

A second remark is that it is clear from our discussion  that  $\cala_k$ generically has multiple  compact boson realizations. Indeed, for every  way of 
splitting $k = p q$, the compact boson at $R^2 = p/q$ gives a realization of  $\cala_k$. 
If $p$ and $q$ are relatively prime, we get the  realization (\ref{freereal}) above, while if\footnote{We use the notation that $(m,n)$ and $[m,n]$ denote the greatest common divisor  resp.  least common multiple of $m$ and $n$.}  $(p,q)=\a >1$, 
the compact boson theory instead realizes the larger algebra $\cala_{k \over \a^2}$ in which  $\cala_k$ is contained. 
Only in the case that $k$ is  prime does   $\cala_{k}$ have a unique compact boson realization. 

\subsection{Representations and characters}
Primary representations of $\cala_k$ are built on states created by vertex operators $:e^{i p_L X_L}:$ which are local with respect to the currents $W^\pm$, leading to 
\be 
p_L = {\n \over \sqrt{k}}, \qquad \n \in \ZZ \label{pLrep}
\ee
Since for $\n = \pm 2 k$ these operators are precisely $W^\pm$, the index $\n$ should be taken in the range
\be
- 2k < \n < 2 k.
\ee
The corresponding $\cala_k$ characters will be denoted as $\chi_{\n,k}$ and   are given by
\be 
\chi_{\n, k }= {1 \over \h} \sum_{n \in \ZZ} q^{( 2 n k + \n)^2 \over 4 k},
\ee
where 
$\h = q^{1\over 24}\prod_{m=1}^\infty (1- q^m)$ is  Dedekind's eta function.
Some of these representations are equivalent due to the identities
\be \chi_{\n, k} = \chi_{-\n, k}=\chi_{\n+ 2 k, k}.\label{Kids}
\ee
Therefore the inequivalent representations have characters $\chi_{\l, k}$ where  $\l$ is restricted to the range $0 \leq \l \leq k$.

The subalgebra relation $\cala_{ k} \subset \cala_{ k/\a^2}$, for  $\a$ a quadratic divisor of $k$, is reflected in the following  branching formula
for the characters:
\be
\chi_{\n,{k\over \a^2}} = \sum_{\x=0}^{\a-1}  \chi_{\a \n + {2  k \x\over \a},  k}.\label{alphaconv}
\ee
\subsection{Modular invariants}\label{Secmodinvs}
Under modular $S$- and $T$-transformations, the characters transform as
\begin{align}
T:  && \chi_{\n, k} (\t+1) =& e^{ 2 i \p \left({\n^2 \over 4 k} - {1 \over 24}\right)}  \chi_{\n, k} (\t)\\
S:  && \chi_{\n, k} \left(- {1 \over \t}\right) =& {1 \over \sqrt{2 k}} \sum_{\n'= 0 }^{2k-1} e^{i \p {\n \n'\over k}} \chi_{\n', k} (\t).\label{STrep}
\end{align}
In the last expression, we should keep in mind that some of the terms on the right hand side are linearly dependent due to (\ref{Kids}). As it stands, the matrix representing $S$ in the basis $\{ \chi_{\l, k},\l =0 ,\ldots , k\}$ is unitary with respect to a nonstandard metric which is diagonal with eigenvalues  $ (1, 2,2, \ldots, 2, 1)$. 
	For this reason we introduce the  rescaled basis functions  $\k_{\l,k}$:
\be
\k_{0,k} \equiv \chi_{0,k}, \qquad \k_{k,k} \equiv \chi_{k,k}, \qquad \k_{i,k} \equiv \sqrt{2} \chi_{i,k} {\ \ \rm for}\ \ 1 \leq i\leq k-1.\label{Lambdas}
\ee
One verifies that in this basis both $S$ and $T$ are represented  by  unitary $(k+1) \times (k+1)$ matrices  
satisfying
\be
S^2 = ( S T )^3 = 1.
\ee
These matrices therefore generate a unitary representation of the modular group $\G \equiv PSL(2, \ZZ )$. 

The most general modular invariant combination of $\cala_k \times \overline{\cala_k}$ characters was found in \cite{Cappelli:1986hf,Cappelli:1987xt} (see also \cite{DiFrancesco:1987gwq} for a useful summary).
From the above realizations in terms of free bosons we know that, for every divisor $\d$ of $k$, we get a modular invariant from a boson at radius ${R  }= {\sqrt{k} \over \d}$: 
\be Z^\d \equiv  Z\left[ {\sqrt{k} \over \d}\right],\label{Zdelta}\ee
 Here, $Z[R]$ is the compact free boson partition function  given by
\be
Z[R] = {1\over |\h|^2}\sum_{w, m \in \ZZ} q^{{1 \over 4}\left({m\over {R}}  + w  {R} \right)^2} \bar q^{{1 \over 4}\left({m\over {R}}  - w  {R} \right)^2}.\label{ZRfree}
\ee

In what follows we will abbreviate the greatest common denominator of $\d$ and $k/\d$ by $\a$:
\be
\a :=  \left(\d, {k\over \d} \right).\label{alphadef}
\ee
Note that $\a$ is a  quadratic  divisor of $k$, $\a^2 |k$.
It is a standard result \cite{DiFrancesco:1987gwq}  that 
 the free boson partition function (\ref{ZRfree}) can be expanded in terms of  characters as 
 \be 
Z^\d  =   \sum_{\n=0}^{2 k-1} \chi_{\n,{k \over \a^2}} \overline{\chi_{\o \, \n,{k \over \a^2} }}
.\label{Zboschars}
\ee
The quantity $\o$ appearing in this expression is determined by $\d$ as follows. 
From (\ref{alphadef}) and Bezout's lemma, there exist integers $r,s$ such that \be r  {\d \over \a} - s  {k \over \a \d} = 1.\label{rsdef}\ee We then define $\o$ as
\be 
\o = \left[ r  {\d \over \a} + s  {k \over \a \d}\right]_{2k\over \a^2} 
\label{omdef}
\ee
Because of (\ref{Kids}),  $\o$ is only determined modulo $2 k/\a^2$, and one easily checks that a different choice of $r$ and $s$ leads to the same $\o$. 
The proof of (\ref{Zboschars}) relies on the property
\be 
\o \left( p {\d \over \a} - q  {k \over \a \d} \right) \equiv  p {\d \over \a} + q  {k \over \a \d}\ \left(\md {2k\over \a^2}\right),
\ee
for arbitrary integers $p,q$.

Let us comment on the properties of the modular invariants $Z^\d$.
We see from (\ref{Zboschars}) that  all they  are all physical modular invariants in the sense that  each character appears with positive integer multiplicity and that  the vacuum  character appears with multiplicity one. The invariant corresponding to $\d =1$ is the diagonal modular invariant of $\cala_k \times \overline{\cala_k}$, while for $\d = \a$, with $\a$ quadratic divisor, we have a nondiagonal modular invariant which is however diagonal with respect to  the extended algebra  $\cala_{k/\a^2} \times \overline{\cala_{k/\a^2}}$.

If $\a >1$, the expression  (\ref{Zboschars}) involves $\cala_{k/\a^2}$ characters  $\chi_{\n, k/\a^2}$ and  still needs to be re-expressed in terms of $\cala_k $ characters $\chi_{\n, k}$ using the  branching formula (\ref{alphaconv}). 
Upon doing so, we can write the expression  (\ref{Zboschars}) for  $Z^\d$ in     terms of a $2k \times 2k$ matrix $M^\d$ 
as follows:
\be
Z^\d = \a \sum_{\n,\n'=0}^{2k -1} \chi_{\n,k} M^\d_{\n, \n'} \overline{ \chi_{\n',k}},\label{Kdecomp}
\ee
Note the somewhat unusual normalization by the factor $\a$ in this definition; this leads to somewhat nicer  properties of the matrices $M^\d$ to be discussed in section \ref{Secnonprim} below.  The explicit component expression of $M^\d$ is
\be 
M^\d_{\n, \n'} =\left\{ \begin{array}{ll} {1\over \a} \d_{[\n'- \o \n]_{2 k\over \a}}&  {\ \rm if \ } \a| \n  { \rm \ and\ }  \a| \n'  \\
	0 &{\ \rm otherwise}  \end{array} \right., \qquad 0 \leq \n, \n'< 2k\label{Mcomp}
\ee
Here, the quantity $\d_{[x]_n}$ is defined to be one if $x\equiv 0\ (\md n)$ and zero otherwise.
For later reference we list some properties of the matrices $M^\d$, which are straightforward to derive:
\begin{itemize}
	\item They are symmetric,
	\be M^\d_{\n, \n'}=M^\d_{\n', \n}.\label{Missymm}\ee
\item Under  exchange of  $\d$ and $\d /k$, which sends $\o \to -\o$, they satisfy
	\be
	M^\d_{0,\n} = M^{k/\d}_{0,\n}, \qquad M^\d_{i,\n} =  M^{k/\d}_{2 k -i ,\n} \qquad {\rm for \ } i = 1, \ldots, 2k-1 .\label{Mssymm}
	\ee
\end{itemize}

We proceed by  converting (\ref{Kdecomp}) to the unitary basis (\ref{Lambdas}), which defines a $(k+1) \times (k+1)$  matrix $I^\d$ through
\be 
Z^\d =\a \sum_{\l, \l'= 0}^k \k_{\l, k} I^\d_{\l \l'} \overline{ \k_{\l ' , k}}.\label{Idef}
\ee
More explicitly, the matrices $I^\d$ are also symmetric and, using (\ref{Mssymm}), are related to  $M^\d$  as
\begin{align}
I^\d_{0,0} =& M^\d_{0,0}, & I^\d_{0,k} =& M^\d_{0,k}, & I^\d_{k,k} =& M^\d_{k,k}\nonu
I^\d_{0,i} =&  \sqrt{2} M^\d_{0,i}, & I^\d_{k,i} =&  \sqrt{2} M^\d_{k,i} &&\nonu
I^\d_{i,j} =& M^\d_{i,j}+ M^{\d/k}_{i,j} & && {\rm for \ }i,j =& 1 , \ldots, k-1 .\label{IitoM}
\end{align}
By construction, the matrices $I^\d$ are modular invariant in the sense that they commute with the $(k+1)$-dimensional unitary matrices representing the action of $S$ and $T$ (see  (\ref{STrep})) in the basis (\ref{Lambdas}).  
	
Summarizing, we have constructed a modular invariant partition function for every divisor $\d$ of $k$. This set of modular invariants is however twofold redundant:  because of T-duality,
 \be 
 Z[R] = Z[1/R],
 \ee
the divisors  $\d $ and ${k \over \d}$ lead to the same modular invariant.
As a check,  using the properties
(\ref{Mssymm}) and (\ref{IitoM}) one sees that the modular invariant matrices indeed satisfy
\be 
I^\d = I ^{k \over \d}.
\ee
A set of independent modular invariant matrices is therefore obtained by restricting the  divisors to the range  $\d^2 \leq k$,
\be 
\{ I^\d,\  \d |k, \ \d^2 \leq k \}.\label{Ibasis}
\ee 
The result of \cite{Cappelli:1986hf,Cappelli:1987xt}  is that (\ref{Ibasis})  furnishes a complete basis  of $\cala_k \times \overline{\cala_k}$ modular invariants. We note that the number of modular invariants is $d(k)/2$, where $d(k)$ is the number of divisors of $k$.

\subsection{Poincar\'e sums}
After these preliminaries we are ready to study the object of interest already introduced   in (\ref{Poincsumintr}).  We define a  `gravity partition function' (some justification for this name will be provided in section \ref{Secbulk}) as the average over the modular group of  the $\cala_k \times \overline{\cala_k}$ vacuum character\footnote{It would be straightforward to generalize this to different `seed' characters, though we will not do so in this work.}, 
 i.e.:
\be 
Z_{\rm grav} (\t, \bar \t) := \caln \sum_{\g \in \G_f \backslash \G} | \chi_{0,k} (\g \t )|^2 \label{Zgravdef}. 
\ee  
Here, $\G = PSL(2, \ZZ)$ is the modular group and $\G_f$ is defined to be the subgroup leaving the vacuum character $| \chi_{0,k}|^2$ invariant. We have included an as yet unspecified normalization factor $\caln$;  summing over the full modular group would lead to an (infinite) overall factor which can be absorbed   $\caln$.  As usual, the elements $\g \in \G$ are represented by $2 \times 2$ matrices and act on $\t$ by fractional linear transformations,
\be 
\g \t := { a \t + b \over c \t + d}, \qquad \g = \left( \begin{array}{cc} a & b\\ c & d \end{array}\right).
\ee 
Modular invariant expressions of the  type (\ref{Zgravdef}) are called  Poincar\'e sums, and in rational CFTs
the sum  contains only a finite number of terms \cite{Castro:2011zq,Meruliya:2021utr}. In our case,  the transformation properties  (\ref{STrep}) imply \cite{Cappelli:1986hf} that $\G_f$ contains the principal congruence subgroup $\G ( 4 k)$   and is therefore  a finite index subgroup of the modular group.

Since  (\ref{Zgravdef}) is modular invariant, it can be expanded in the basis (\ref{Ibasis}): 
\be
Z_{\rm grav} = \sum_{\d, \d^2 \leq k} c_\d Z^\d \label{average}
\ee
We would like to interpret  $Z_{\rm grav}$ as an averaged partition function over the ensemble of rational torus CFTs, with the coefficients $c_\d$ 
playing the role of ensemble weights. This is of course only sensible if the $c_\d$ are positive. If so,  we will fix the  normalization $\caln$ in (\ref{Zgravdef}) such that the weights add up to one,
\be 
\sum_{\d , \d^2 \leq k } {c_\d  }  =1.\label{normcs}
\ee

Before going on it is illuminating   to work out the Poincar\'e sum (\ref{Zgravdef}) explicitly in some simple examples. First let us consider the case $k=4$. Trying out combinations of the generators $S$ and $T$, one finds that the factor group $\G_f \backslash \G$ contains  24 elements, namely 
\be
\G_f \backslash \G= \{ 1, S T^{i_1} , S T^2 S T^{i_2}, S T^4 S, S T^8 S, S T^{12} S ; \   i_1 = 0, \ldots ,15, \ i_2 = 0,\ldots , 3 \}
\ee
 There are two modular invariants corresponding to the divisors $\d = 1,2$. Performing the sum in (\ref{Zgravdef}), one finds the linear combination
\be
Z_{\rm grav } = \caln ( 4 Z^{ 1} +  Z^{2} ).
\ee
Imposing (\ref{normcs}) we find the normalization factor and 
 the ensemble weights in (\ref{average}):
\bea
\caln &=& 1/5,\nonu
c_{ 1}&=& {4 \over 5}, \qquad c_{2}= {1 \over 5}. \label{Ncsk4}
\eea

Next let us discuss the case $k=6$. The factor group $\G_f \backslash \G$ consists of  48 elements:
\bea
\G_f \backslash \G &=& \left\{ 1, S T^{i_1} , S T^2 S T^{i_2}, S T^3 S T^{i_3}, S T^4 S T^{i_4}, S T^6 S T^{i_5}, S T^8 S T^{i_6}, S T^{12} S ;\right.\nonu
	&& \left. \   i_1 = 0, \ldots ,23, \ i_2 = 0,\ldots , 5, \ i_3 = 0,\ldots , 7, \ i_4 = 0,1 , 2, \ i_5 = 0,1 , \ i_6 = 0,1 , 2 \right\}\nonumber
\eea
There are again two modular invariants corresponding to the divisors $\d = 1,2$, and  performing the sum in (\ref{Zgravdef}) we find that the ensemble weights are in this case equal: 
\be
Z_{\rm grav } = 4 \caln (  Z^{ 1} +  Z^{2} )
\ee
leading to 
\bea
\caln &=& 1/8, \label{Nkis6} \\
c_{ 1}&=&  c_{ 2}= {1 \over 2}. \label{cskis6}
\eea
These examples illustrate  a general feature which we will derive below: when $\cala_k$ is primitive (i.e. when $k$ is square-free), all the weights in the ensemble average are equal. In the non-primitive case, when $k$ has nontrivial square divisors, the weights are in general different.

 For general level $k$, a formula for the weights $c_\d$ can be derived  by viewing (\ref{average}) as an equality between modular matrices in the space of characters as in (\ref{Idef}). 
Using modular invariance of the matrices $I^\d$, one obtains the relation  \cite{Castro:2011zq,Meruliya:2021utr}

\be 
\sum_{\d', (\d')^2 \leq k} \tr \left(I^\d I^{\d '}\right) \left(\d',{k \over \d'}\right) c_{\d'}={\caln [\G : \G_f] } \tr ( I^{\d} X^{\rm vac} ).\label{cdeltagen1}
\ee
Here, $[\G : \G_f]$ is the index of $\G_f$ in $\G$
and $X^{\rm vac}$ is the matrix representing the `seed' partition function in the space of characters. In our case of interest we simply have
$ X^{\rm vac}_{ \l \l'} = \d_{\l,0}\d_{\l' ,0}$. Using also   that in our normalization $I^\d_{00} =  (\d, k/\d)^{-1}$, (\ref{cdeltagen1}) 
reduces to
\be 
\sum_{\d', (\d')^2 \leq k}D_{\d,\d'} c_{\d'}={\caln [\G : \G_f] },\label{cdeltagen}
\ee
where  $D$ is a matrix with components
\be 
D_{\d ,\d'}:= \left(\d, {k\over \d}\right) \left(\d', {k\over \d'}\right) \tr \left( I^\d I^{\d'}\right). \label{Dcomps}
\ee
 To obtain the ensemble weights $c_\d$
it looks at first sight like we have to invert the square matrix $D$ of dimension $d(k)/2$. However,  we shall see that the problem is drastically simplified because of  additional structure on the space of modular invariants.
In the next two sections we will work  this out in more detail  for the primitive and non-primitive cases.

\section{Primitive case}\label{Secprim}

The primitive case, when $k$ is square-free, is by far the simplest.  None of the modular invariants of $\cala_k \times \overline{\cala_k}$  is then associated  to an extended algebra. We will presently prove that in this case {\em all the ensemble weights $c_\d$ in  
	(\ref{average}) are equal.} We will see that the fact that all   modular invariants are on the same footing is a consequence of the property that the matrices $I^\d$ form a group under multiplication \cite{Cappelli:1986hf}.  

To elucidate this group property, it is convenient to change notation and  label modular invariants,   not by the divisor $\d$, but by   the associated quantity $\o$  defined in (\ref{omdef}). We  note that, in the primitive case, for every divisor $\d$ of $k$, we have $\a= (\d , k/\d) =1$. 
From (\ref{rsdef}) and (\ref{omdef})  we see that every $\o$ is a root of unity modulo $4k$,
\be 
\o^2 \equiv 1\ (\md 4k),
\ee 
 and therefore belongs to the multiplicative group
\be 
\tilde G^{1} = \{  [\o ]_{ 2k};   \o^2 \equiv 1\  (\md 4 k) \}.\label{G1def}
\ee
Here the notation  $[x]_n$ means the conjugacy class of $x$ modulo $n$. 
In (\ref{G1def}) the $\o$'s are considered modulo $2k$ due to (\ref{Kids}). 
This is consistent\footnote{One can show that, when $k$ is square-free, one could in (\ref{G1def}) just as well consider roots of unity modulo $2k$, but this will no longer be true for general $k$.} since the   roots of unity modulo $ 4 k$ always come in pairs  $\o$ and $\o + 2 k$.
The map between divisors $\d$ and elements of $\tilde G^1$ is one-to-one since
 the cardinality $|\tilde G^{(1)}|$ equals the number of divisors of $k$, namely $2^{\O (k )}$, where $\O (k)$ is the number of distinct primes entering in the prime decomposition of $k$.
So far we have not yet taken into account the fact that  T-duality identifies  the modular invariants associated  to
$\o$ and $-\o$, and that  inequivalent modular invariants are  therefore 
 in one-to-one correspondence with elements of the quotient group
\be 
G^{1} \equiv  \tilde G^{1}/\{\pm 1\}.
 \ee

 Instead of labelling the modular invariant matrices as $I^\d$ as in (\ref{Ibasis}), we will label them from now on as $I^1_\o$, with $\o \in G^1$.
A crucial   property is that these matrices form a $(k+1)$-dimensional representation $R^{1}$ of $G^{1}$:
\be
I^{1}_\o I^{1}_{\o'} = I^{1}_{\o \o'}. \label{G1mult}
\ee
This property is a special case of the theorem proven in Appendix \ref{Appmult}, but it can also be seen from the component form of 
the $I^{1}_\o$. There is a natural action of the group $G^1$ on the quotient space $\ZZ_{2 k}/\{ \pm 1 \}$
 (the quotient identifies the conjugacy classes $[x]_{2k}$ and $[-x]_{2k}$),  on which $\o$ acts by multiplication. 
It's straightforward to verify from (\ref{Mcomp}, \ref{IitoM}) that the $I^1_\o$ are simply the matrices representing this action on  the elements $\{ [0], \ldots, [k]\}$.

We remark that the resulting representation  $R^1$ 
is always reducible and contains the trivial representation at least twice. 
This can be seen from the properties
\bea 
\o \l &\equiv& 0\ (\md 2k) \Leftrightarrow \l \equiv 0 (\md 2 k)\\
\o \l &\equiv& k\ (\md 2k) \Leftrightarrow \l \equiv k (\md 2 k) \label{singlets}
\eea
for all $ \o \in G^{(1)}$, which follow from $(\o, 2k) =1$ and multiplying both sides by $\o$.	
This  implies that  the $I^{1}_\o$ have the block-diagonal structure 
\be 
(I^{1}_\o)_{0\l} = (I^{1}_\o)_{\l 0}= \d_{\l,0} , \qquad  (I^{1}_\o)_{k\l }=  (I^{1}_\o)_{\l k }= \d_{\l,k}
\ee
and the representation contains at least two singlets.

Now we turn to the formula (\ref{cdeltagen}) for the ensemble weights. In the primitive case it reduces to
\be 
\sum_{\o' \in G^{1} } D_{\o \o'}  c_{\o '}={\caln [\G : \G_f] } , \label{csprimform}
\ee
where the matrix elements (\ref{Dcomps}) of  $D$ are 
\be 
D_{\o ,\o'} = \tr I^{1}_{\o} I^{1}_{\o'}=  \tr  I^{1}_{\o\o'} = \chi^{1} (\o \o').
\ee
Here, $ \chi^{1} (\o )= \tr I^{1} (\o )$ are the characters of the representation $R^{1}$. 
In particular, the sum of the elements in each row of $D$ is the same, namely
\be 
\sum_{\o'} D_{\o \o'} =  \sum_{\o'} \chi^{1} (\o \o')=   \sum_{\tilde \o}  \chi^1 (\tilde \o ) = |G^{1}| N_{\rm id}^{1}, \label{Deigenvect}
\ee
where $N_{\rm id}^{1}$ is the number of times the trivial representation appears in the $k+1$-dimensional representation $R^1$ (from the above  we know that $N_{\rm id}^{1} \geq 2$).
The relation (\ref{Deigenvect}) means  that $(1,1, \ldots, 1)$ is an eigenvector of $D$, and  we see that (\ref{csprimform}) is solved by taking  all $c_\o$ to be equal and given by
\be 
c_\o = {1\over |G^{1}| }= 2^{2 - \O(k)},\label{csprim}
\ee
while the normalization factor satisfies
\be 
\caln [\G : \G_f] = {  N_{\rm id}^{1}  }.\label{Ngen}
\ee

Let us illustrate these properties in the simple example where $k=6$. The two modular invariants are associated to the elements of
\be
G^1 = \{ [1]_{12}, [5]_{12} \}.
\ee
 The  corresponding modular invariant matrices are 
\be 
I^{1}_1 = {\bf 1}_{7 \times 7}, \qquad I^{1}_5 ={\small \left( \begin{array}{ccccccc} 
1&0&0&0&0&0&0 \\
0&0&0&0&0&1&0 \\
0&0&1&0&0&0&0 \\
0&0&0&1&0&0&0 \\
0&0&0&0&1&0&0 \\
0&1&0&0&0&0&0 \\
0&0&0&0&0&0&1
\end{array}\right)}
\ee
These  realize the $G^{1}$ multiplication law, in particular $\left(I^{1}_5\right)^2 = I^{1}_1$, and form the representation $R^1$. By diagonalizing $I^{1}_5$ one sees that
\be 
N_{\rm id}^{1} =6,
\ee
so that (\ref{Ngen}) is consistent with  our earlier  brute-force determination   of $ [\G : \G_f]=48 $ and  $\caln = 1/8$ in (\ref{Nkis6}). That calculation also found the weights to be equal (\ref{cskis6}) in agreement with  the general result (\ref{csprim}). 
\section{Non-primitive  case}\label{Secnonprim}
We now turn to the non-primitive case, where the level $k$ has nontrivial square divisors. We will see that the space of modular invariants, while no longer a group, still possesses the structure of a semigroup. This will allow us to drastically simplify the equation (\ref{average}) for the ensemble weights.

For each  square divisor $\a$   such that $\a^2 |k$  (including $\a =1$) we consider the subset of divisors $\d$ of $k$
such that $( \d, k/\d ) = \a$. Similar to our discussion of the primitive case, where $\a=1$ was the only square divisor, one finds that those divisors $\d$ are in one-to-one correspondence with the elements of the multiplicative group
\be 
\tilde G^{\a} \equiv \left\{  [\o]_{{2 k \over  \a^2}} ; \o^2 \equiv 1 \ \left(\md {{4 k \over \a^2}} \right) \right\}.
\ee
To account for identifications under T-duality we once again consider the quotient groups
\be 
G^{\a} \equiv  \tilde G^{\a}/\{\pm 1\}.
\ee
The cardinality of these abelian groups is $| G^{\a} |=2^{ \O (k/\a^2) -1}$.
In this way we see that the $\cala_k \times \overline{\cala_k}$ modular invariants are in one-to-one correspondence with the elements of the union
\be
\calm = \bigcup_{\a, \a^2 |k} G^{\a} .
 \ee
We will presently see that  there is a natural multiplication law for elements belonging to different $G^{\a}$ which gives $\calm$  the structure of a semigroup.
 
We start by observing that, if $\a$ and $\b$ are square divisors of $k$, then so is their least common multiple $[\a, \b]$. It is straightforward to prove this by decomposing of $\a,\b$ and $k$ into prime factors.
 This suggests to define an operation $\bullet$ on $\calm_k$ which multiplies  elements
 $\o \in G^{\a}$ and $\tilde \o \in G^{\b} $ to yield an element of $ G^{[\a, \b]}$
 as follows:
 \be
 \o \bullet  \tilde  \o := \left[\o  \tilde  \o \right]_{{ 2 k \over [\a,\b]^2 }}, \qquad \qquad  (\o \in G^{\a},\tilde \o \in G^{\b}).
 \ee
To check that the result indeed belongs to $G^{[\a,\b ]}$ one verifies that
 \be	 	
\left(\o   \tilde \o\right)^2 \equiv 1 \  \left(\md{{ 4 k \over [\a,\b]^2 }}\right) \qquad \qquad  (\o \in G^{\a},\tilde \o \in G^{\b}).
	\ee
The $\bullet$ operation gives $\calm$ the structure of a semigroup. It is in addition  commutative, associative and has an identity element (namely $[1]_{2k} \in G^1$) so more precisely  $(\calm, \bullet)$ is an abelian monoid.

Now we consider the modular invariant matrices (\ref{Idef}) which we henceforth relabel as $I^{\a}_{\o}$ for $\o \in G^\a$. For a fixed square divisor $\a$, they can be shown to form a $(k+1)$-dimensional representation of $G^{\a}$  which we will  refer to as $R^{\a}$. 
What is more, they also furnish a representation of the semigroup $(\calm, \bullet)$ under matrix multiplication:
\be 
I^{\a}_{\o} I^{\b}_{\tilde \o} = I^{[ \a, \b ]}_{\o \bullet\tilde  \o}
\qquad \qquad  (\o \in G^{\a},\tilde \o \in G^{\b}).\label{Ismultnonprim}\ee
The proof of this property is given in Appendix \ref{Appmult}.

Having uncovered this structure, let us take a closer look at the formula (\ref{average}) for the ensemble weights, which we rewrite as
\be 
\sum_{\b; \b^2 |k}\  \sum_{\tilde \o \in G^{\b }} D_{\o, \tilde \o }c_{\tilde \o}^\b ={\caln [\G : \G_f] } \qquad  {\rm for \ } \o \in G^{\a}.\label{csnonprim}
\ee 
Since the right-hand side doesn't depend on $\o$, we make the ansatz that the  coefficients  $c^\a_{ \o}$ depend only on $\a$:
\be
c_{\o}^\a = c^\a, \qquad \forall \o \in G^{\a }.\label{csame}
\ee
We now show that this ansatz is justified, as the matrix $D$ maps vectors of this type into each other.
Its matrix elements are, using (\ref{Ismultnonprim})
\bea 
D_{\o, \tilde  \o} &=& \a\b\, \tr I^{\a}_{\o} I^{\b}_{\tilde \o} \\
&=& \a\b\, \chi^{[\a, \b]} ( \o \bullet \tilde \o)  \qquad \qquad  (\o \in G^{\a},\tilde \o \in G^{\b}),
\eea
where $\chi^{\a } (\o ) = \tr I^{\a}_{\o} $ is the character of the representation $R^{\a}$.
We work out the sum
\bea 
\sum_{\tilde \o \in G^{\b }} D_{\o, \tilde \o} &=& 
\a\b \sum_{\tilde \o \in G^{\b }}  \chi^{[\a, \b]} ( \o\bullet \tilde \o) \qquad  ( \o \in G^{\a})\label{superprop1}\\
&=&\a\b N_{\rm id}^{[\a,\b]} {|G^{\b}|\over |G^{[\a,\b]}|},\label{superprop}
\eea
where $ N_{\rm id}^{\a}$ is the number of times the trivial representation appears in the representation $R^{\a}$ of $G^{\a}$. The second line of (\ref{superprop}) follows from the following\\
{\bf Property: }  if $\b$ and $\a$ are square divisors
of $k$ and $\b | \a $, then the map 
\be 
G^{\b} \to G^{\a}: \qquad \o \mapsto  \left[ \o \right]_{2 k \over \a^2}\label{mapGs}
\ee
is $N$-to-one, where \be N = {|G^{\b}|\over |G^{\a}|} = 2^{\O( k/\b^2) - \O( k/\a^2)}.\label{Ntoone}\ee \\
From this property, which we prove in Appendix \ref{Appmaps}, we see that when performing the sum in (\ref{superprop1}) the argument of $ \chi^{[\a, \b]} $ cycles through all the elements of $G^{[\a, \b]}$ precisely $|G^{\b}| / |G^{[\a,\b]}|$ times, and the result (\ref{superprop}) follows.

Substituting (\ref{csame},\ref{superprop}) in (\ref{csnonprim}) we obtain a formula for the ensemble weights
\be
c^\a_{\o} = \caln [\G : \G_f] \sum_\b  \left(d^{-1}\right)_{\a,\b} \qquad \forall \o \in G^\a, \label{csfinal}
\ee
where $d$ is the matrix with elements
\be
d_{\a, \b} =\a\b  N_{\rm id}^{[\a,\b]} {|G^{\b}|\over |G^{[\a,\b]}|}.
\ee
The upshot of this somewhat lengthy analysis is that we have simplified the original problem of inverting the matrix $D$ in (\ref{average}), whose size is half the number of divisors of $k$,  to that of inverting the matrix $d$ whose size is the number of square divisors of $k$. It will be interesting to see if future efforts can produce an analytic formula for the  weights (\ref{csfinal}), or a proof that they are always positive (we did not find any counterexamples).

Let us illustrate the formula (\ref{csfinal}) in  the simple subclass where $k = 4 p$ with $p$ an odd prime.
There are two quadratic divisors, $1$ and $2$, and when  $p>1$ there are three independent modular invariants. The groups $G^{1,2}$ are
\be 
G^1= \{ [1]_{8p},[2 p \pm 1]_{8p}\},\qquad G^2= \{ [1]_{2p}\},
\ee
where the $\pm$ sign is to be chosen such that $4 | (p\pm 1)$. The corresponding modular invariant matrices satisfy the semigroup multiplication rules
\be
\big( I^1_{2 p \pm 1}\big)^2 = I^1_1,\qquad  I^1_{2 p \pm 1} I^2_1 = I^2_1, \qquad \big( I^2_1\big)^2 = I^2_1,
\ee
and $I^1_1$ is the identity. One verifies that
\be 
N^1_{\rm id} =\half \tr \left( I^1_1 +  I^1_{2 p \pm 1} \right)= {5\over 2}(p+1),\qquad  N^2_{\rm id} = \tr  I^2_1 = p+1.
\ee
In the first equality we used that $\tr I^1_{2 p \pm 1}=p+4$, which  follows from the fact that the only elements of $\ZZ_{8p}/\{ \pm 1 \}$ left invariant by multiplication by $2 p \pm 1$ are the multiples of 4 and $p$. 
Applying (\ref{csfinal}) and imposing the normalization condition (\ref{normcs}) we find
\bea
\caln |\G : \G_f| &=& {4 \over 3}(p+1)\nonu
c^1_1 &=& c^1_{2 p \pm 1} = {4 \over 9} ,\qquad c^2_1 ={1 \over 9}.
\eea
In the special case that $p=1, k=4$, there are only two modular invariants,
\be 
G^1= \{ [1]_{8}\},\qquad G^2= \{ [1]_{2}\},
\ee
and we similarly find $N^1_{\rm id}=5, N^2_{\rm id} =2$ leading to
\bea
\caln |\G : \G_f| &=& {24 \over 5},\\
c^1_1 &=&  {4 \over 5} ,\qquad c^2_1 ={1 \over 5}, 
\eea
 in agreement with our earlier brute-force calculation  of $|\G : \G_f|=24$ and $\caln$ in  (\ref{Ncsk4}).
We should remark that the solution to (\ref{csnonprim}) is not always of the form $c^\a_\o = { c  \over \a^2}$ as one might be tempted to guess from this example.

\section{Bulk interpretation as an $U(1)^2$ exotic gravity}\label{Secbulk} 
In this section we discuss the tentative bulk dual interpretation of the ensemble average over $\cala_k \times \overline \cala_k$ rational torus CFTs encountered in the previous sections. In analogy with  Narain duality \cite{Afkhami-Jeddi:2020ezh,Maloney:2020nni,Perez:2020klz,Dymarsky:2020pzc,Datta:2021ftn,Benjamin:2021wzr,Ashwinkumar:2021kav,Dong:2021wot,Collier:2021rsn,Benjamin:2021ygh} and earlier examples of rational ensemble holography \cite{Castro:2011zq,Meruliya:2021utr,Meruliya:2021lul}, 
the bulk dual would be an exotic gravity theory described by a Chern-Simons action supplemented by a prescription to sum over certain topologies in the path integral.
In the case of interest, we have two compact $U(1)$ Chern-Simons theories at levels $k$ and $-k$, and we will review some of the standard dictionary with rational torus CFT on the boundary \cite{Moore:1989yh,Elitzur:1989nr}. In particular, we recall  the role of solitonic field configurations in the bulk which lead to the chiral currents $W^\pm$ of the $\cala_k$ algebra.  
We will then consider the bulk partition function on thermal AdS$_3$ and show that including the sum over soliton sectors leads to  the
 $\cala_k \times \overline \cala_k$ vacuum character
\be
Z_{\rm thermal AdS}  =|\chi_{0,k}|^2.\label{vacPF}
\ee
Assuming the sum over topologies to comprise those of the $PSL(2, \ZZ )$ family of Euclidean geometries appearing in semiclassical gravity \cite{Maldacena:1998bw} 
 one then obtains a bulk justification of the Poincar\'{e} sum (\ref{Poincsumintr}).

\subsection{Solitons and $\cala_k$}
We consider two abelian Chern-Simons theories\footnote{We should stress that we consider here fully decoupled Chern-Simons theories, as in the original Chern-Simons-rational CFT literature \cite{Witten:1988hf,Moore:1989yh,Elitzur:1989nr}. A different theory, where the Chern-Simons fields are coupled through a boundary term $-{k\over 2 \p} \int_{\d \calm} A \wedge \bar A$, has been argued to be much more trivial  in \cite{Witten:2003ya,Maloney:2020nni}.} with opposite values of the level:
\be 
S = S[A] - S[\tilde A], \qquad S [A] = - {k \over 2 \p} \int_\calm A dA .\label{SCS}
\ee
The gauge group consists of two compact $U(1)$'s, 
\be 
A  \to A +  G^{-1} d G, \qquad |G| =1,
\ee
and the level $k$ should be quantized to be an integer for the path integral to be well-defined.

\begin{figure}\label{FigDirac}
 	\begin{center}
 		\includegraphics[height=120pt]{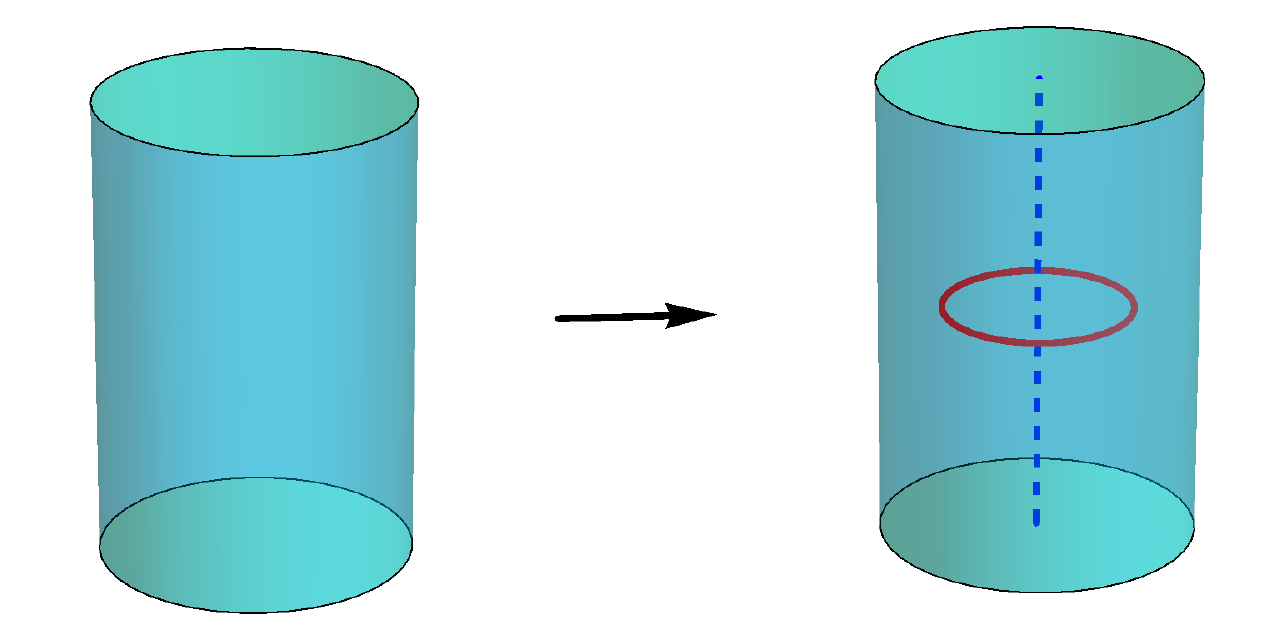}
 		\put(-20,60){\red{ $\calc$}}
 		\put(-35,22){\blue{ $\call$}}
 	\end{center}
 	\caption{A large gauge transformation produces a Dirac string singularity along the dotted line $\call$, which is however invisible as any Wilson loop encircling it along $\calc$ is trivial.}
 \end{figure}

We consider this theory on global AdS$_3$, or more precisely its conformal compactification which has the topology of a solid cylinder, i.e. $\calm = \RR \times D$ with $D$ the disk. 
As stressed in \cite{Moore:1989yh,Elitzur:1989nr} 
an important role in identifying  the $\cala_k \times \overline \cala_k$ symmetry of the theory  is played by topological 
sectors which appear due to the gauge group not being simply connected. Let us illustrate this in more detail, focusing on one $U(1)$ factor for simplicity, the other one proceeding analogously. 
Consider the effect of a large gauge transformation where, as we go around the angular direction $\f$ on the disk, the group element goes around $U(1)$ circle multiple  times.  The gauge parameter is of the form $G = e^{i n \f}$, where $n \in \ZZ$ and sends 
 $ A \to A + n d\f$. Using $dd\f = 2 \p \d( r) dr \wedge d \f$, performing such a large gauge transformation in the path integral introduces a  Wilson line of charge $2kn$: 
 \be 
W_{2 kn} [\call] =e^{2 i k n \int_{\call} A},\label{Wilsonlarge}
 	\ee
where $\call = \RR\times \{ 0 \}$ is a line in the middle of the solid cylinder.
	We should therefore view these  Wilson lines as part of  our pure gauge theory, rather than describing external matter. In the presence of (\ref{Wilsonlarge}), the gauge field has a Dirac string singularity along $\call$,   see Figure \ref{FigDirac}, which is however invisible to  physical matter fields: when encircling the Dirac string along a closed curve $\calc$, the wavefunction of a  charge $q $  matter field ($q \in \ZZ)$ picks up a phase 
\be 
W_q [\calc]=  e^{ i q \int_{\calc} A} = 1.
\ee
 
 These topological sectors are responsible for extending  the  chiral algebra from  a $u(1)$ current algebra to $\cala_k$, in particular the currents $W^\pm$  correspond to $n = \pm 1$.
 To see this, we make the standard reduction of $U(1)$ Chern-Simons theory to a chiral boson on the boundary cylinder \cite{Moore:1989yh,Elitzur:1989nr}. 
We start by  imposing the following boundary condition on the  cylinder $\RR \times S^1$:
\be 
{A_-}_{| \RR \times S^1} =0.
\ee
where $x_\pm \equiv \f \pm t$.
Upon partially integrating, the time component $A_t$ becomes a Lagrange multiplier for the  first-class constraint
\be 
F_{ij} =0,
\ee
where $ i,j = 1,2$ are components on the disk $D$.
Solving the constraint as
\be 
A_i = \pa_i \L,
\ee
where $\L \sim \L + 2\p $, the action, including the boundary term coming from the above partial integration, reduces to the Floreanini-Jackiw action \cite{Floreanini:1987as} for a  chiral scalar  on the boundary,
\be 
S = -{k \over \p}\int_{\RR \times S^1} d^2 x \pa_\f \L \pa_- \L \label{Schirbos}
\ee
The conserved energy and $\f$-momentum  are
\be 
H =  P =  {k \over 2 \p} \int_{S^1} d\f (\pa_\f \L)^2.
\ee

The solitonic sectors discussed above correspond to  winding solutions where
\be
\L = n \f, \qquad n \in \ZZ.
\ee
These carry the conserved charges
\be
\D := \half (H + P)  = k n^2, \qquad \bar \D := \half (H - P)  = 0,
\ee  
and therefore, for $n = \pm 1$, they carry the quantum numbers of the chiral currents $W^\pm$. A similar analysis of the second $U(1)$ theory leads to the antiholomorphic currents $\overline{ W^\pm}$.
The analysis of \cite{Moore:1989yh,Elitzur:1989nr} shows that canonical quantization yields the Hilbert space $\calh_{\RR \times D}$  on the solid cylinder to be the vacuum module of $\cala_k \times \overline \cala_k$.

\subsection{Path integral  on the solid torus}
Next we discuss the path integral of the Chern-Simons theory (\ref{SCS}) on thermal AdS$_3$ whose conformal compactification is a solid torus, $\calm = S^1 \times D$. From our discussion of the Hilbert space   on the solid cylinder one would expect that the result should be the $\cala_k \times \overline \cala_k$ vacuum character,
\be
 Z_{ S^1 \times D} \stackrel{?}{=} \tr_{\calh_{\RR \times D}} q^\D \bar q^{\bar \D} =
 |\chi_{0,k}|^2,\label{Zsoltorus}
 \ee
 where $q = e^{2 \p i \t}$ and $\t$ the modular parameter of the boundary torus.
A small extension of standard results \cite{Porrati:2019knx},   including the contributions from soliton sectors, shows that this is indeed the case, and we include it here for completeness. As above we make use of the reformulation as a boundary chiral boson theory, though presumably one could also give a more direct bulk computation along the lines of \cite{Porrati:2019knx,Maloney:2020nni}.

We focus  on the path integral for the first $U(1)$ field $A$.  It can be shown \cite{Elitzur:1989nr} that it reduces to a  path integral over the boundary field $\L$ with action (\ref{Schirbos}) without any  Jacobian factors.  To improve the convergence properties we define the integral through analytic continuation to imaginary time
\be 
t \to i t_E. \label{Eucltime}
\ee
We now want to compute the path integral
\be 
Z^{U(1)}_{\calm} = \int {[D\L] \over V_{\rm gauge}} e^{ - S_E [\L]},\qquad
S_E[ \L ]= {k \over \p} \int_{\d \calm} \L '\pa_{\bar w} \L,\label{chiralPI}
\ee
where $w := \f + i t_E$. The division by the volume $V_{\rm gauge}$ arises because the action has a restricted gauge invariance
\be 
\L \to \L + \a( t_E  ).\label{Lambdagauge}
\ee
The boundary $\d \calm$ in (\ref{chiralPI}) is taken to a torus with complex structure parameter $\t$, as reflected in the periodicities
\be 
( t_E, \f ) \sim ( t_E, \f + 2 \p) \sim ( t_E + 2 \p \t_2, \f + 2 \p \t_1 ).
\ee
As argued in the previous subsection, we should allow winding sectors around the $\f$-circle. It seems natural to also  allow winding sectors around the Euclidean time circle, though as we will presently see these  do not contribute to the path integral. In the sector with winding numbers $n$ and $m$ around the respective cycles the  boundary conditions on the field $\L$ are
\bea 
\L (t_E, \f+ 2\p)&=& \L (t_E, \f) + 2 \p n,\\
\L (t_E + 2 \p \t_2, \f+ 2\p \t_1) &=&\L (t_E, \f ) + 2 \p m. \label{bcLambda}
\eea

The classical solution obeying these boundary conditions is 
\be 
\L_{\rm cl} = n  \f +{m  - n \t_1 \over \t_2} t_E,\label{Lclass}
\ee
on which the action takes the value
\be 
S_{E, \rm cl} = - 2\pi i k n ( n  \t-m) .\label{Sclass}
\ee
We note from (\ref{Lclass}) that the winding number $m$ can be removed by a large gauge transformation of the type (\ref{Lambdagauge}) and from (\ref{Sclass}) that it does not contribute to $e^{ - S_{E, \rm cl} }$. The sum over winding sectors labelled by $m$ therefore contributes an $\t$-independent (infinite) normalization factor, which cancels against a similar factor in $V_{\rm gauge}$.

We  expand the field around the classical solution as
\be
\L = \L_{\rm cl} + k^{-\half} \psi,
\ee
where $\psi$ is single-valued on the boundary torus, i.e. it satisfies (\ref{bcLambda}) with $n=m=0$. Performing the gaussian integral\footnote{We note that, thanks to the imaginary time continuation (\ref{Eucltime}), the operator $- \pa_\f\pa_{\bar w} $ has positive real part and the Gaussian integral is well-defined.}  over $\psi$ and  summing over the winding sectors leads to 
\be 
Z^{U(1)}_{\calm} = \left( \det\,' (- \pa_\f\pa_{\bar w} ) \right)^{-\half}\sum_{\n \in \ZZ} q^{k n^2}.\label{Gaussint}
\ee
Here the prime on the determinant that it excludes the $\f$-independent zero modes, which cancel with $V_{\rm gauge}$ in (\ref{chiralPI}). The regularized functional determinant 
can be computed with standard methods, see e.g. Appendix A of \cite{Porrati:2019knx}, and equals 
$\h^{-1}$. Therefore the path integral leads to the $\cala_k$ vacuum character
\be 
Z^{U(1)}_{\calm} = {1\over \h}\sum_{\n \in \ZZ} q^{k n^2} = \chi_{0,k}.
\ee
Similarly  the path integral in the  $U(1)_{-k}$ theory  leads to $\overline {\chi_{0,k}}$ and combining the two we arrive at (\ref{vacPF}).

\subsection{The sum over topologies}
Since the boundary description of the $U(1)_k \times U(1)_{-k}$ Cherns-Simons theory includes a stress tensor, constructed as a bilinear in the $u(1)$ current, $T = :J^2:$, the theory can  be thought of as including gravity  in some sense. As such, a nonperturbative definition would involve summing over topologies in the path integral.
However, since the Virasoro central charge is one, the gravity sector is strongly coupled and we have no semiclassical control over the  topologies we should include, and the prescription is necessarily ad hoc. In the case of a genus-one boundary, a natural family of topologies to include is the 
$PSL(2,\ZZ)$ family of Euclidean BTZ solutions encountered in semiclassical gravity \cite{Maldacena:1998bw}. With this assumption, the bulk path integral leads to the Poincar\'{e} sum (\ref{Poincsumintr}) which was our starting point.

\section{Comment on multiple boundaries and wormholes}\label{Secmultibound}
So far we have argued that the  path integral of an exotic bulk $U(1)_k\times U(1)_{-k}$  theory over manifolds with a genus one boundary yields  an ensemble averaged  partition function over rational torus CFTs. To   extend these observations  into a bona fide holographic duality
one should in principle show that all observables computed  in the bulk  reduce to   CFT observables in the {\em same} weighted ensemble average of theories.  Examples of such  observables are path integrals in the presence of multiple and/or higher genus  boundaries. 

Here, we will comment on the path integral with multiple genus one boundaries, which was already studied in the context of ensemble holography for rational CFTs in \cite{Meruliya:2021utr}. The bulk path integral should contain wormhole-like contributions from geometries which connect several boundary components and  whose value is severely constrained by consistency. We want to illustrate  how these consistency conditions simplify in ensemble averages where
all CFTs have equal weights, as is the case in the  primitive $\cala_k$ theories. In particular, we will derive an analytic expression for the wormhole contributions when the ensemble contains two  CFTs    with equal ensemble weights. This happens for our exotic $U(1)_k\times U(1)_{-k}$ theory at $k = p q$, where $p$ and $q$ are nonequal primes.

Let us denote by $ Z_1 (\t), Z_2 (\t)$  the torus partition functions of the two CFTs, so that the gravity path integral on a manifold with a single torus boundary is
\be
Z^{(1)}_{\rm grav}(\t_1) =\half \Big(  Z_1 (\t) +  Z_2 (\t) \Big)\label{Z1}
\ee 
In the presence of $b$ boundary components with modular parameters $\t_i, i = 1, \ldots , b$, we will denote by $Z^{(b)}_{\rm grav}(\t_1, \ldots , \t_b) $ the total bulk path integral  and by $Z^{(b)}_{\rm conn}(\t_1, \ldots , \t_b) $ its connected part, coming (for $b>1$) from wormhole-like geometries which connect all the boundaries.
The total  path integral should  arise from combining the various connected contributions, for example for $b =1,2,3$ we  have
\bea
Z^{(1)}_{\rm grav}(\t_1) &=& Z^{(1)}_{\rm conn}(\t_1)\label{Z1conn}\\
Z^{(2)}_{\rm grav}(\t_1,\t_2 ) &=& Z^{(1)}_{\rm conn} (\t_1)  Z^{(1)}_{\rm conn} (\t_2) +  Z^{(2)}_{\rm conn} (\t_1, \t_2) \label{Z2grav}\nonu
Z^{(3)}_{\rm grav}(\t_1,\t_2,\t_3 ) &=& Z^{(1)}_{\rm conn} (\t_1)  Z^{(1)}_{\rm conn} (\t_2)  Z^{(1)}_{\rm conn} (\t_3)\nonu
& + & Z^{(2)}_{\rm conn} (\t_1, \t_2)  Z^{(1)}_{\rm conn} (\t_3) +  Z^{(2)}_{\rm conn} (\t_1, \t_3)  Z^{(1)}_{\rm conn} (\t_2)  +  Z^{(2)}_{\rm conn} (\t_2, \t_3)  Z^{(1)}_{\rm conn} (\t_1)\nonu
& + & Z^{(3)}_{\rm conn} (\t_1, \t_2, \t_3)\label{Z3grav}
\eea
 
On the other hand, for consistency of the ensemble interpretation, the gravity  path integral with  $b$ boundaries should  also be equal to an ensemble average
with equal weights, namely
\be
Z^{(b)}_{\rm grav}(\t_1, \ldots , \t_b) = \half \Big( Z_1(\t_1) \ldots Z_1(\t_b) + Z_2(\t_1) \ldots Z_2(\t_b)  \Big)\label{ZB}
\ee
Comparing with (\ref{Z3grav}) and its generalization to include more boundary components  one finds that the $b$-boundary wormhole contribution should be of the form
\be
 Z^{(b)}_{\rm conn} (\t_1, \ldots , \t_b) = w_b   \Big(  Z_1 (\t_1) -  Z_2 (\t_1) \Big) \ldots \Big(  Z_1 (\t_b) -  Z_2 (\t_b) \Big), \qquad {\rm for \ } b>1,\label{ZconnB}
\ee
with the first few coefficients $w_b$ given by
\be 
w_2 = {1 \over 4}, \qquad w_3 =0, \qquad w_4 = -{1 \over 8}.\label{lowds}
\ee

A recursion relation for the coefficients $w_b$ was derived for arbitrary  ensemble weights  in \cite{Meruliya:2021utr}. In our special case of equal ensemble weights we can do better and derive an analytic expression for the $w_b$ as follows. 
The generalization of (\ref{Z3grav}) to arbitrary boundaries can be expressed as a standard relation between connected and non-connected  generating functions. It is of the schematic form, suppressing the dependence on the modular parameters,
\be 
\ln \left( 1 +  \sum_{b=1}^\infty {k^b Z^{(b)}_{\rm grav}\over b! } \right)=
 \sum_{b=1}^\infty {k^b Z^{(b)}_{\rm conn}\over b! },
\ee
Substituting (\ref{ZB}) in the left-hand  side we derive 
\be
 \sum_{b=1}^\infty {k^b Z^{(b)}_{\rm conn}\over b! } = {k \over 2} (Z_1 + Z_2) + \ln \cosh {k \over 2} (Z_1 - Z_2)
 \ee
 Taylor expanding the right-hand side, one finds the results  (\ref{Z1conn}) and (\ref{ZconnB}), where the coefficients $w_b$ are given by
 \be
 w_b = \left\{ \begin{array} {ll} { (2^b - 1) B_b \over b} & {\rm for\ } b \ {\rm even } \\
 0	& {\rm for\ } b \ {\rm odd } \end{array} \right.
\ee
Here, $B_n$ is the $n$-th Bernoulli number. One checks that this agrees with (\ref{lowds}) for $b = 2,3,4$.

\section{Outlook}
In this note we have studied simple examples of ensemble holography which can be seen as rational CFT versions of the Narain holography proposed in  \cite{Afkhami-Jeddi:2020ezh,Maloney:2020nni}. It is our hope that these can help shed light on some of the  puzzles surrounding 
the  meaning and generality of averaged holography. 
  
  One aspect of ensemble holography which is readily illustrated in our examples is that it is a feature of path integrals in {\em low energy effective theories which are not UV complete}. This is embodied in the fact that our bulk theory only contained massless gauge fields, leading to a spectrum on global AdS$_3$ consisting of only the vacuum module, \be 
  Z_{\rm thermal\ AdS_3}= |\chi_{0,k}|^2.\ee The sum over topologies  leads
  to a modular invariant answer (\ref{Poincsumintr}) which is however agnostic as to the precise UV completion and ends up sampling all UV complete CFTs with the relevant chiral algebra.
   Once we give more information about UV in the form of the matter content, the ensemble of CFTs sampled in the result  narrows down \cite{Maloney:2020nni}. The extreme case of this is if we would add matter fields in the bulk which fill out a modular invariant spectrum, for example leading to the diagonal invariant  
   \be 
Z_{\rm thermal\ AdS_3}= |\chi_{0,k}|^2 + 2 \sum_{\l = 1}^{k-1}  |\chi_{\l,k}|^2 + |\chi_{k,k}|^2 .\ee
   Then the sum over bulk topologies and the corresponding modular average is trivial and doesn't lead to an ensemble average. Similarly, other UV-complete theories such as the tensionless string  on AdS$_3$ \cite{Eberhardt:2018ouy,Eberhardt:2020bgq} and the limit of W$_N$ minimal models dual to 3D vasiliev theory \cite{Gaberdiel:2012uj},  have  modular invariant spectra on global AdS$_3$.
   
   We end by listing some generalizations and open problems.
\begin{itemize}
	\item It should be straightforward yet interesting to generalize our examples to rational CFTs involving several bosons on an even integral lattice.
	\item In sections \ref{Secprim} and \ref{Secnonprim} we saw that the formula for the ensemble weights drastically simplifies thanks to an additional (semi)group  structure on the space of modular invariants. It would be interesting explore such structures and their consequences for ensemble holography  in rational Virasoro or Kac-Moody CFTs. Also, it would be satisfying to give an interpretation of the number-theoretic formula (\ref{csfinal}) for the ensemble weights in terms of CFT data. 
	\item As we illustrated in section \ref{Secmultibound}, the wormhole-like contributions in the presence of multiple boundaries
	are fixed by consistency and fully calculable in some cases. It would be very interesting to have bulk derivation of these contributions, perhaps along the lines of the results \cite{Cotler:2020ugk,Cotler:2020hgz} in pure gravity.
	\item In order to give a semiclassical justification for the  geometries included in the path integral, it would be useful to have simple examples of averaged holography where the gravity sector is weakly coupled, i.e. where the Virasoro central charge is parametrically large (rather than equal to one as in the current work). By a spectral flow operation  on the $\cala_k$ algebra one can obtain a closely related algebra where Virasoro central charge is large and negative, and which we argued in our previous work \cite{Raeymaekers:2020gtz} to govern a nonstandard semiclassical limit of pure gravity.  It seems likely that the current results can be reinterpreted as a version of ensemble holography for pure gravity in this limit, and we hope to report on this in the near future.
\end{itemize}
 \section*{Acknowledgement}
 This work was supported
by the Grant Agency of the Czech Republic under the grant EXPRO 20-25775X.
\begin{appendix}
\section{Some  number-theoretic lemmas}\label{Apparithm}
In this Appendix we prove some  properties which follow from elementary number theory (see e.g. \cite{niven1991introduction}) and which are needed in section \ref{Secnonprim}.
\subsection{Semigroup property of modular matrices}\label{Appmult}
We would like to prove  the multiplication property (\ref{Ismultnonprim}): for $\a, \b$ square divisors of $k$ and for any $\o \in G^\a$ and $\tilde \o \in G^\b$, 
\be 
I^{\a}_{\o} I^{\b}_{\tilde \o} = I^{[ \a, \b ]}_{\o \bullet\tilde  \o},
\label{multIs}
\ee
where $\o \bullet\tilde  \o = [ \o\tilde  \o ]_{2 k/ [\a,\b]^2}$. Since, by definition,  
 $I^\a_\o = I^\a_{[\o]_{2k/\a^2}}$, we can write (\ref{multIs}) more simply as
 \be 
 I^{\a}_{\o} I^{\b}_{\tilde \o} = I^{[ \a, \b ]}_{\o\tilde  \o} \qquad (\o \in G^\a, \tilde \o \in G^\b).\label{multIs2}
 \ee
We will first show that the property holds for the matrices $M^\a_\o$ defined in (\ref{Kdecomp}), i.e. \be M^{\a}_{\o} M^{\b}_{\tilde \o} = M^{[ \a, \b ]}_{\o\tilde  \o}\qquad (\o \in G^\a, \tilde \o \in G^\b).\ee If $\a | \n $ and $\b|\n'$ then the component expression (\ref{Mcomp}) gives   
\be 
\left(M^{\a}_{\o} M^{\b}_{\tilde \o}\right)_{\n \n'} = (\a \b)^{-1} \times \left|\left\{ x \in \ZZ_{2 k} , \left\{ \begin{array}{c} x \equiv \o \n (\md {2 k \over \a}) \\  x \equiv \tilde \o \n' (\md {2 k \over \b})
		\end{array}\right.  \right\} \right|\label{multMcomp}
\ee	
The generalization of the Chinese remainder theorem to non-coprime moduli  tells us that the simultaneous congruence in the brackets on the right-hand side has solutions if and only if
\be
\o \n \equiv \tilde \o \n' \left(\md \left( {2k \over \a},  {2k \over \b} \right) \right) .\label{condsols}	 
\ee
Note that the modulus in this equation can be written as $\left( {2k \over \a},  {2k \over \b} \right) = { 2k \over [\a,\b]}$.  One shows that, since $\a | \n $ and $\b|\n'$, (\ref{condsols}) implies that $[\a, \b]$ divides both $\n$ and $\n'$. Using  that $ (\o, {2k \over [\a,\b]^2})=1$, we can multiply both sides of (\ref{condsols}) by $\o$ to find the equivalent statement
	\be
	 \n \equiv \o \tilde \o \n' \left(\md { 2k \over [\a,\b]}. \right) 	 
	\ee
	If this is satisfied, the solution is unique modulo  $ \left[ {2k \over \a},  {2k \over \b} \right] = {2k [\a, \b] \over \a \b}$. The number of solutions of the congruence in $\ZZ_{2k}$ then equals ${\a \b \over [\a, \b ]}$. Equation 
	(\ref{multMcomp}) then becomes
\bea 	\left(M^{\a}_{\o} M^{\b}_{\tilde \o}\right)_{\n \n'} &=&
  \left\{ \begin{array}{c l} {1 \over [\a, \b]}  \d_{ [\n - \o \tilde \o \n'] _ { 2k / [\a,\b]}} & {\rm if \ } [\a, \b] | \n \ {\rm and \ } [\a, \b] | \n' \\
  	0 & {\rm otherwise}
 \end{array} \right.\\
&:=& 	\left( M^{[ \a, \b ]}_{\o\tilde  \o}\right)_{\n \n'}
  \eea
	
From this property one derives  the analogous multiplication property  (\ref{multIs2}) for the matrices $I^\a_\o$ using the relations (\ref{IitoM}) and (\ref{Mssymm}). Let us illustrate how this works for the $0 i$ component of  (\ref{multIs2}), where $i = 1, \ldots , k-1$:
\bea
\left(I^{\a}_{\o} I^{\b}_{\tilde \o}\right)_{0 i} &=& \big(I^{\a}_{\o}\big)_{00} 
\big(I^{\b}_{\tilde \o}\big)_{0 i}+\big(I^{\a}_{\o}\big)_{0j} 
\big(I^{\b}_{\tilde \o}\big)_{ji}+\big(I^{\a}_{\o}\big)_{0k} 
 \big(I^{\b}_{\tilde \o}\big)_{k i}, \qquad \qquad j = 1, \ldots, k-1\nonu
 &=& \sqrt{2} \left(\big(M^{\a}_{\o}\big)_{00} 
 \big(M^{\b}_{\tilde \o}\big)_{0 i}+\big(M^{\a}_{\o}\big)_{0j} 
\left( \big(M^{\b}_{\tilde \o}\big)_{ji} +  \big(M^{\b}_{- \tilde \o}\big)_{ji}\right)+\big(M^{\a}_{\o}\big)_{0k} 
 \big(M^{\b}_{\tilde \o}\big)_{k i}\right)\nonu
&=&  \sqrt{2} \left(\big(M^{\a}_{\o}\big)_{00} 
 \big(M^{\b}_{\tilde \o}\big)_{0 i}+\big(M^{\a}_{\o}\big)_{0j} 
 \big(M^{\b}_{\tilde \o}\big)_{ji} + \big(M^{\a}_{\o}\big)_{0 2k-j}  \big(M^{\b}_{ \tilde \o}\big)_{2k - ji}+\big(M^{\a}_{\o}\big)_{0k} 
 \big(M^{\b}_{\tilde \o}\big)_{k i}\right)\nonu
 &=& \sqrt{2} 	\left( M^{[ \a, \b ]}_{\o\tilde  \o}\right)_{0i}\nonu
 &=& \left( I^{[ \a, \b ]}_{\o\tilde  \o}\right)_{0i}.
 \nonumber
 \eea
 The equality for the other components is proven in a similar manner. 
 \subsection{Maps between the groups  $G^\a$}\label{Appmaps}
 Next we turn to the proof of the property surrounding eqs. (\ref{mapGs},\ref{Ntoone}).
 Let us consider two quadratic divisors $\a$ and $\b$ of $k$, such that $\b | \a$. Then we have a map  from the group  $\tilde G^\b$ to the group $\tilde G^\a$ provided by
 \be 
  \tilde G^\b \to \tilde G^\a, \qquad \o \mapsto  [\o ]_{2k /\a^2}.\label{mapba}
 	\ee
We want to show that this map
 is $N$-to-one, where \be N = {|\tilde G^{(\b)}|\over |\tilde G^{(\a)}|} = 2^{\O( k/\b^2) - \O( k/\a^2)}.\ee 
 Since the map (\ref{mapba}) commutes with multiplication by $-1$, the property also holds for the  map between the quotient groups   $G^\a = \tilde G^\a/ \{ \pm 1 \}$ and  $G^\b = \tilde G^\b/ \{ \pm 1 \}$.
 
To prove the property, we observe that by redefining $k \to k/b^2$ and  $\a \to [\a, \b]$ we can restrict our attention to the case $\b =1$.  Let us choose an arbitrary element
$\tilde \o \in G^\a$.  We need to show that the pre-image  of $\tilde \o$ under the map (\ref{mapba}) consists
of precisely $N$ elements, independent of the chosen $\tilde \o$. We recall from section \ref{Secmodinvs}
that $\tilde \o$ is uniquely defined by a divisor $\tilde \d$ of $k$ satisfying $(\tilde \d, k/\tilde \d) =\a$.
More precisely, given a pair of  numbers $\tilde r, \tilde s$ from Bezout's lemma such
that
\be
\tilde r \tilde \d - \tilde s k/\tilde \d = \a,\label{Beztilde}
\ee
 $\tilde \o$ can be written as
 \be 
 \tilde \o = \left[ \tilde r {\tilde \d \over \a}+ \tilde s { k \over  \a \tilde \d }\right]_{2 k \over \a^2}.\label{tildeoBB}
 \ee
The numbers $\tilde r, \tilde s$  are not unique but the class $\tilde \o$ does not depend on which ones we choose. Our property will follow from choosing $\tilde r, \tilde s$
in a judicious way.
 
 To do so, we first decompose  $\a$ and $k$ in prime factors, 
 \bea
 \a &=& p_1^{a_1} \ldots   p_n^{a_n}\\ 
 k &=&   p_{1}^{2 a_{1}} \ldots   p_t^{2 a_t} p_{t+1}^{2 a_{t+1} + c_{t+1} } \ldots p_n^{2 a_n + c_n}  p_{n+1}^{d_{n+1}} \ldots   p_{m}^{d_m}
 \eea
  Here, $t$ is by definition smaller than or equal to $n$, and where $t<n$    the numbers $c_i, i = t+1 , \ldots n$ are defined to be strictly positive.
Some reflection shows that the integers $m,n$ and $t$ have the interpretation \be m = \O (k), \qquad n = \O(\a ), \qquad t =  \O (k)- \O(k / \a^2).\ee

 The divisor $\tilde \d$ is of the form
 \be 
 \tilde \d =  p_{1}^{ a_{1}} \ldots   p_t^{ a_t} p_{N+1}^{ a_{t+1} + \tilde b_{t+1} c_{t+1} } \ldots p_n^{ a_n + \tilde  b_n c_n}  p_{n+1}^{\tilde b_{n+1} d_{n+1}} \ldots   p_{m}^{\tilde b_m d_m}.
 \ee
It is specified by the numbers  $\tilde b_{t+1}, \ldots \tilde b_n$ which take the value zero or one. 
 
 On the other hand, the elements of $\tilde G^1$ are in one-to-one correspondence with divisors $\d$ with $(\d, k/\d)=1$. These are of the general form
 \be 
\d =   p_{1}^{2 b_1 a_{1}} \ldots   p_t^{2 b_t a_t} p_{t+1}^{b_{t+1}(2 a_{t+1} + c_{t+1}) } \ldots p_n^{b_n(2 a_n + c_n)}  p_{n+1}^{b_{n+1} d_{n+1}}  \ldots   p_{m}^{b_m d_m},
\ee
 where $b_1, \ldots b_m$ are either zero or one. 
 
 To our specific $\tilde \d$, specified by $\tilde b_{t+1}, \ldots \tilde b_n$, we now associate $N = 2^t$ different divisors $\d$ of the form
 \be 
   \d =   p_{1}^{2 b_1 a_{1}} \ldots   p_t^{2 b_t a_t} p_{t+1}^{\tilde b_{t+1}(2 a_{t+1} + c_{t+1}) } \ldots p_n^{\tilde b_n(2 a_n + c_n)}  p_{n+1}^{\tilde b_{n+1} d_{n+1}}  \ldots   p_{m}^{\tilde b_m d_m}. \label{deltadeltat}
   \ee
   Here, each $b_i, i = 1, \ldots t$ can take the value zero or one, leading indeed to $2^t$ possibilities.
   
  Our property follows if we can   show that, for each choice of $\d$ of the form (\ref{deltadeltat}), the associated $\o_\d$ satisfies
  \be 
  \left[ \o_\d \right]_{2k \over \a^2} = \tilde \o.
  \ee
  First, we rewrite $\d$ as
  \be 
  \d = { \tilde \d \over \a} V
  \ee
  where
  \be 
 V =  p_{1}^{2 b_1 a_{1}} \ldots   p_t^{2 b_t a_t} p_{t+1}^{\tilde 2 b_{t+1} a_{t+1} } \ldots p_n^{2 \tilde b_n 2 a_n }
  \ee
  We note that
  \be 
  V | \a^2. \label{divasq}
  \ee
  Now we pick some $r,s$ such that
  \bea 
 1 &=&  r \d  - s { k \over \d} \\
  &=&  r V  { \tilde \d \over \a} - s { \a^2 \over  V} { k \over \a \tilde \d}.
  \eea
We then choose $\tilde r, \tilde s$ satisfying (\ref{Beztilde})  as follows
 \be 
 \tilde r =  r V, \qquad \tilde s = s { \a^2 \over  V} ,
 \ee
 the latter being integer thanks to (\ref{divasq}). With this choice equation   (\ref{tildeoBB}) becomes
 \bea 
 \tilde \o &=&  \left[  r \d  + s { k \over \d}\right]_{2k \over \a^2}\\
 &=&  \left[  \o_\d \right]_{2k \over \a^2},
 \eea
 for each of the $N$ choices for $\d$.
\end{appendix}                                                                                                                                                                       
\bibliographystyle{ytphys}
\bibliography{refsolitons}
\end{document}